\documentclass[aps,prd,twocolumn,a4paper]{revtex4}
\usepackage{ulem}
\usepackage{color}
\usepackage{float}
\usepackage{graphicx}
\usepackage{epsfig}
\usepackage{epstopdf}
\usepackage{amsmath}
\usepackage{subfig}
\usepackage[dvipsnames]{xcolor}
\newcommand{\be}{\begin{equation}}
\newcommand{\ee}{\end{equation}}
\newcommand{\ba}{\begin{eqnarray}}
\newcommand{\ea}{\end{eqnarray}}
\newcommand{\nn}{\nonumber\\}

\usepackage{graphicx}
\usepackage{color}
\begin{document}
\title{Relativistic second order dissipative hydrodynamics from effective fugacity quasi particle model}
\author{Sukanya Mitra}
\email{sukanya.mitra10@gmail.co.in}
\affiliation{National Superconducting Cyclotron Laboratory, Michigan State University, East Lansing, Michigan 48824, USA}

\begin{abstract}
In this work a second order relativistic viscous hydrodynamic model has been presented based on the effective fugacity quasi-particle model (EQPM).
The hydro model has been derived from the effective relativistic second-order transport equation under EQPM for a multi-particle (two component) 
system and solving it in Grad's 14 moment method. The EQPM model describes the strongly interacting thermal system of QCD interactions through its 
fugacity parameters extracted from an updated lattice equations of state. The proper time evolution of temperature and pressure anisotropy is observed 
to be affected significantly due to the inclusion of EQPM model compared to an ideal system.
\\
\\
{\bf  Keywords}: Effective fugacity quasi particle model, second order relativistic hydrodynamics\\ 

\end{abstract}
\maketitle

\section{Introduction}

In last few decades, during the flourishing era of investigating deconfined quark-gluon plasma (QGP) through heavy-ion collision experiments,
relativistic hydrodynamics has been proved to be the most trusted tool to describe the system properties \cite{Landau,Blaizot,Bjorken,Jeon,Kolb,Song}. Hydrodynamics provides the collective
behavior of the system through evolution equations of the macroscopic state variables. However, since hydrodynamics only deals with bulk properties 
of the systems and do not care about the single particle distributions, the microscopic dynamics of the system is not included in this theory.
But a microscopic theory is essential in its own merit since the input parameters of these evolution equations crucially depend upon particle
interactions and underlying dynamics. Many particle kinetic theory has been so far served well for this purpose by describing the macroscopic
thermodynamic quantities and transport coefficients in terms of single particle distribution functions and system interactions \cite{Danielewicz,Arnold,Hosoya,Heiselberg,Dusling,Chen}. Techniques
that consistently bridge between these two theories have been quite successful in describing the matter created in heavy ion collision
experiments.

In the due course, a few earlier attempts were made with ideal hydrodynamics considering the system as a weakly interacting gas which seems to offer first handedly a sensible description 
of the data. However, a closer inspection to some of the bulk observables such as multiplicity, radial or elliptic flow reveals the need to include dissipative hydrodynamics 
in order to describe the space-time evolution of the system \cite{Luzum}. However, the first order or the Navier-Stokes theory results in parabolic equations of motion of the thermodynamic state 
variables leading to severe causality violation problem. This crisis requires the introduction of a second order theory which provides hyperbolic equations of motion resulting
in finite time scale for the thermodynamic flows to dissipate. This second order hydrodynamic theory which is famously known as Muller-Israel-Stewart theory, despite of having
some issues with numerical stability, has been served the study of dynamical evolution of the system quite authentically \cite{MIS,Muronga1,Romatschke,Wiedemann,Calzetta,Monnai,Jaiswal,Denicol0,Denicol3,Denicol1,Strickland}.

In this work, the second order relativistic hydro equations have been derived and solved for a 1+1 boost invariant system including the effective fugacity quasiparticle model (EQPM) \cite{Chandra}.
The hydro equations have been derived from relativistic transport equation of covariant kinetic theory by solving them applying Grad's 14 moment method within a viscous medium.
The crux of the work lies into appropriate modeling of the equilibrium, isotropic momentum distribution of gluons and quarks in the hot QCD medium under EQPM scheme.
The EQPM scheme, with a recent (2+1)-flavor lattice QCD equation of state (LEOS) \cite{Bazavov} at physical quark masses, have been exploited in the present manuscript.

The basic idea of quasiparticle models are to describe the hot QCD equations of state (EOSs) in terms of non-interacting or weakly interacting effective gluons and quarks.
Previously, second order relativistic hydro have been studied using different quasiparticle theories, especially within the scope of effective mass models \cite{Tinti,Alqahtani,Alqahtani1}.
The novelty of EQPM model resides with the fact that the hot QCD medium effects present in the QCD EOSs (either computed within improved perturbative QCD (pQCD) or lattice 
QCD simulations) can be implemented with a single temperature dependent fugacity parameter for each parton. The resulting particle distribution can describe all the 
thermodynamic quantities consistently without introducing a temperature dependent mass and the dispersion relation of the underlying theory remains linear. The 
mean field force term induced by the system's collective behavior is simply implemented by the fugacity parameters and conserves the particle current and energy-momentum
tensor perfectly. Further, the model has been recently extended to be used with finite quark chemical potential as well \cite{Mitra1}. So, the extracted hydrodynamic
equations with this model and their solutions are expected to implement a realistic equation of state of the medium within the formalism of a consistently
developed covariant kinetic theory.

The manuscript is organized as follows. Section II deals with the covariant formalism of the EQPM model itself, first order theory using Chapman-Enskog
technique to extract first order transport coefficients and the Grad's moment method to derive the second order hydrodynamic equations as well as its 
solution for a 1+1 boost invariant system.  Section III displays the results involving proper time evolution of the concerned thermodynamic quantities and
section-IV ends the manuscript with a conclusion and possible outlook.

\section{Formalism} 

\subsection{Covariant theory for EQPM model}

As mentioned earlier, the basic idea of EQPM model is to map the strongly interacting hot QCD effects into  a medium consisting of practically non-interacting quasi-quarks 
and quasi-gluons through a parameter called effective fugacity $z_{q/g}$. Including this parameter, the equilibrium single particle distribution function for this 
quasi-partons belonging to $k^{th}$ species ($k=q,g$) takes the following form \cite{Mitra1,Mitra2},

\begin{equation}
f^0_{k}=\frac{1}{z_{k}^{-1} \exp\big\{\frac{ E_{p_{k}}}{T}-\frac{\mu_{B_{k}}}{T}\big\}\mp 1}~,
\label{dist1}
\end{equation}
with $E_{p_{k}}$ as the energy of each bare particle and $\mu_{Bk}$ as baryon chemical potential for $k^{th}$ species. $T$ is the temperature of the system at local thermal
equilibrium. We define the four momentum of the quasi-particles as $p_{k}^{\mu}=(\omega_{p_{k}},|\vec{p_{k}}|)$, where the three momenta $|\vec{p_{k}}|$ is the same
as of the bare particles but the energy of each quasi-particle differs from that of bare particles by a dispersion relation,

\begin{equation}
 \omega_{p_{k}}=E_{p_{k}}+\Delta_k,~~~~~~~~~~~~~~\Delta_k=T^2 \partial_{T}ln{z_{k}}~.
 \label{dis}
 \end{equation}
For the current purpose we are considering a massless QGP for which bare particle energy equals with its momentum,  $E_{p_{k}}=|\vec{p_{k}}|$.
The equilibrium distribution function of quasi-partons from  Eq.(\ref{dist1}) can be written alternatively as the following,
\begin{equation}
 f^0_{k}=\frac{1}{\exp\big\{\frac{p_k^{\mu}u_{\mu}}{T}-\frac{\mu_{k}}{T}\big\}\mp 1}~,
 \label{dist2} 
\end{equation}
with $u^{\mu}$ as the hydrodynamic four-velocity and $\mu_k=\mu_{B_k}+\Delta_k+Tlnz_k$ denoting $exp(\mu_k/T)$ as the total effective fugacity.

From \cite{Mitra3} it can be recalled that the particle four flow and energy-momentum tensor under EQPM have the following form,

\begin{eqnarray}
N^{\mu}(x)=&&\sum_{k=1}^{N}\nu_k\int\frac{d^{3}|\vec{p_k}|}{(2\pi)^3{{\omega_{p}}_k}}{p_k^{\mu}}f_k(x,p_k)\nonumber\\
           +&&\sum_{k=1}^{N}\nu_k\Delta_k\int\frac{d^{3}|\vec{{p_k}}|}{(2\pi)^3{\omega_{{p}_{k}}}}\frac{\langle{{{p}}_{k}^{\mu}}\rangle}
           {|\vec{{p_k}}|}f_k(x,{p_k})~,           
\label{PFF}\\
T^{\mu\nu}(x)=&&\sum_{k=1}^{N}\nu_k\int\frac{d^{3}|\vec{{p_k}}|}{(2\pi)^3{\omega_{p_k}}}{{p_k}^{\mu}}{{p_k}^{\nu}}f_k(x,{p_k})\nonumber\\
 +&& \sum_{k=1}^{N} \nu_k\Delta_k\int\frac{d^{3}|\vec{{p_k}}|}{(2\pi)^3{\omega_{p_k}}}\frac{\langle\langle{{p_k}^{\mu}}{{p_k}^{\nu}}\rangle\rangle} {|\vec{{p_k}}|}f_k(x,{p_k})~.
\label{EMT}
\end{eqnarray}
Here $\langle{{p}^{\mu}}\rangle=\Delta^{\mu\nu}{p}_{\nu}$ and 
${\langle\langle{{p}^{\mu}}{{p}^{\nu}}\rangle\rangle}=\frac{1}{2}\big\{ \Delta^{\mu\alpha}\Delta^{\nu\beta}+\Delta^{\mu\beta}\Delta^{\nu\alpha} \big\}{p}_{\alpha}{p}_{\beta}$ 
are the irreducible tensors of rank one and two respectively, with $\Delta^{\mu\nu}=g^{\mu\nu}-u^{\mu}u^{\nu}$
as the projection operator. Throughout the analysis the metric $g^{\mu\nu}$ has taken to be $g^{\mu\nu}=(1,-1,-1,-1)$.

From above equations the expressions of particle number density, energy density and pressure take the following forms respectively,

\begin{eqnarray}
 n(x)=&&\sum_{k=1}^{N}\nu_k\int\frac{d^{3}|\vec{p_k}|}{(2\pi)^3} f_k(x,p_k)~,
 \label{Nd}\\
 \epsilon(x)=&&\sum_{k=1}^{N}\nu_k\int\frac{d^{3}|\vec{p_k}|}{(2\pi)^3} \omega_{p_k} f_k(x,p_k)~,
 \label{Ed}\\
 P(x)=&&\frac{1}{3}\sum_{k=1}^{N}\nu_k\int\frac{d^{3}|\vec{p_k}|}{(2\pi)^3} |\vec{p_k}| f_k(x,p_k)~,
 \label{Pr}
\end{eqnarray}
with $f_k(x,p_k)$ is the single particle momentum distribution belonging to $k^{th}$ species, that is a function of space-time coordinate $x$ 
and particle momenta $p_k^{\mu}$. $\nu_k$ is the corresponding degeneracy factor.

The relativistic transport equation of a single quasi-particle distribution function for a process $p_k+p_l\rightarrow p'_k+p'_l$, including 
the mean field force term resulting from collective excitation of quasi-quarks and quasi-gluons is given by,

\begin{eqnarray}
 \frac{1}{\omega_{p_k}} p_{k}^{\mu}\partial_{\mu}f_k(x,{p_k})+\vec{F_k}\cdot\vec{\nabla}_{p_{k}} f_{k}= \sum_{l=1}^{N}C_{kl}[f_{k},f_{l}]~,\nonumber\\
 ~~~~~~~~~~[k=1,....,N]
 \label{RTE1}
\end{eqnarray}
with $\vec{F_k}$ as the mean field force on $k^{th}$ particle and $C_{kl}$ as the collision integral respectively defined by,
\begin{equation}
 F_{k}^{i}=-\partial_{\mu}\big\{\Delta_k \ u^{\mu} u^{i} \big\}~,
 \label{Force}
 \end{equation}
 and
 \begin{eqnarray}
 &&C_{kl}[f_{k},f_{l}]=\frac{\nu_{l}}{\omega_{p_k}} \int d\Gamma_{{p}^{}_{l}} d\Gamma_{{p}'_{k}}  d\Gamma_{{p}'_{l}}  W
          \nn &&\times  [f_{k}({p}'_{k}) f_{l}({p}'_{l}) 
          \{1\pm f_{k}({p}_{k})\}\{1\pm f_{l}({p}_{l})\}\nn&&-
          f_{k}({p}_{k})f_{l}({p}_{l})\{1\pm f_{k}({p}'_{k})\}\{1\pm f_{l}({p}'_{l})\}]~.
\label{coll1}     
\end{eqnarray}
Here, $d\Gamma_{{p}_{i}}=\frac{d^3 \vec {{p}}_{i} }{(2\pi)^3 \omega_i}$ denotes the phase space factor and 
$W=\frac{1}{2}(\pi)^4 \delta ^4 (p_{k}+p_{l}-p'_{k}-p'_{l})\langle|M_{k+l\rightarrow k'+l'}|^2\rangle$ is the interaction cross section for the corresponding dynamical processes
with $M$ as the scattering amplitudes.
The conservation of particle current $\partial_{\mu}N^{\mu}=0$ and energy-momentum tensor $\partial_{\mu}T^{\mu\nu}=0$ can be trivially realized from Eq.(\ref{PFF}) and (\ref{EMT})
with the help of Eq.(\ref{RTE1}).

Before concluding this section the expressions for shear and bulk viscous pressure tensor under EQPM can be derived respectively as \cite{Mitra3},
\begin{eqnarray}
\langle \Pi^{\mu\nu}\rangle=&&\sum_{k=1}^{N}  \nu_k\int d\Gamma_{p_{k}}\langle{{p_k}^{\mu}}{{p_k}^{\nu}}\rangle f^{0}_k(1\pm f^{0}_k)\phi_k\nonumber\\
             +&&\sum_{k=1}^{N}\nu_k\Delta_k \int d\Gamma_{p_{k}} \frac{\langle{{p_k}^{\mu}}{{p_k}^{\nu}}\rangle}{|\vec{{p_k}}|} f^{0}_k(1\pm f^{0}_k)\phi_k~,
 \label{Pimn}\\
 \Pi=-&&T^2\sum_{k=1}^{N}  \nu_k\int d\Gamma_{p_{k}} \hat{Q}_k f^{0}_k(1\pm f^{0}_k)\phi_k\nonumber\\
             -&&T^2\sum_{k=1}^{N}\nu_k\Delta_k \int d\Gamma_{p_{k}} \frac{\hat{Q}_k}{|\vec{{p_k}}|} f^{0}_k(1\pm f^{0}_k)\phi_k~,
 \label{bulkPi}
 \end{eqnarray}
 with $\hat{Q}_k=\frac{\tau_k^2}{T}\gamma+\tau_k\gamma_k+\frac{1}{3}\frac{\mid\vec{p_k}\mid^2}{T^2}$. 
 $\tau_k=\frac{p^{\mu}_k u_{\mu}}{T}=\frac{\omega_{p_{k}}}{T}$ is the $k^{th}$ particle energy scaled by temperature $T$ in a comoving frame ($u^{\mu}=1,0,0,0$) and
 the coefficients $\gamma$ and $\gamma_k$ are given in the appendix. The notation $\langle\rangle$ denotes the traceless irreducible tensor of rank 2 defined as 
$\langle A_{\mu} B_{\nu} \rangle =\Delta_{\mu\nu\alpha\beta}A^{\alpha}B^{\beta}=\frac{1}{2}\{\Delta_{\mu\alpha}\Delta_{\nu\beta}+\Delta_{\nu\alpha}\Delta_{\mu\beta}-\frac{2}{3}\Delta_{\mu\nu}\Delta_{\alpha\beta}\}A^{\alpha}B^{\beta}$.

\subsection{First order theory}

In this section the first order transport coefficients, particularly shear viscous coefficient will be estimated from covariant multi-component kinetic
theory using a technique called Chapman-Enskog method. The method has been detaily discussed in \cite{Mitra1,Mitra4,Mitra5,Mitra6}.
This is basically an iterative technique, where from the known lower order distribution function the unknown next order can be determined by successive 
approximation. Using this method to first order, Eq.(\ref{RTE1}) turns out to be,

\begin{equation}
 p_{k}^{\mu}\partial_{\mu}f_{k}^{0}+\omega_{p_k}{F_{k}^{i}}\frac{\partial f^{0}_k}{\partial p^{i}_{k}}
 =-\sum_{l=1}^{N}{{\cal L}_{kl}}[\phi_{k}],~~~~~~[k=1,\cdots,N]~.
 \label{RTE2}
\end{equation}
The next to leading order particle distribution function $f_{k}=f_{k}^0+f_{k}^0(1\pm f_{k}^0)\phi_k$ is expressed in terms of deviation in particle's momentum
distribution $\phi_k$. The linearized collision term becomes,
\begin{eqnarray}
 {\cal L}_{kl}[\phi_k]=&&\nu_{l}\int d\Gamma_{p_l} d\Gamma_{p'_k} d\Gamma_{p'_l}f_{k}^{0} f_{l}^{0}\{1 \pm f_k^{'0}\} \{1\pm f_{l}^{'0}\}\nonumber\\
&&[\phi_{k}+\phi_{l}-\phi^{'}_{k}-\phi^{'}_{l}] W(p_{k},p_{l}|p^{'}_{k},p^{'}_{l})~,
\label{coll2} 
\end{eqnarray}
where the primed quantities are the distribution functions and deviation functions with final state momenta.

Operating the derivatives of the left hand side of Eq.(\ref{RTE2}) on the equilibrium distribution function from (\ref{dist2}), we obtain a number
of thermodynamic forces as follows \cite{Mitra1},

\begin{eqnarray}
&&Q_{k} X - \langle p_{k}^{\mu} p_{k}^{\nu} \rangle \langle X_{\mu\nu}\rangle + \langle p_{k} ^{\nu}\rangle \{(p_{k}.u)-h_{k}\}X_{q\nu}
\nn&&+ \langle p_{k}^{\nu} \rangle \sum_{a=1}^{N'-1}(q_{ak}-x_{a})X_{a\nu}  = -\frac{T}{f_k^0(1\pm f_k^0)}\sum_{l=1}^{N}{{\cal L}_{kl}}[\phi_{k}]~,\nonumber\\
 \label{RTE3}
\end{eqnarray}
with,
\begin{eqnarray}
\label{eq-T20}
X=&&\partial \cdot u ~,\\
X_{q\mu}=&&[\frac{\partial_{\mu}T}{T}-\frac{\partial_{\mu}P}{nh}]~,\\
\label{eq-T21}
X_{k\mu}=&&[(\partial_{\mu}\mu_{k})_{P,T}-\frac{h_{k}}{nh}\partial_{\mu}P]~,\\
\label{eq-T22}
\langle X_{\mu\nu}\rangle=&& \langle \nabla _{\mu} u_{\nu} \rangle ~.        
\label{RTE4}
\end{eqnarray}

Here $Q_{k}=\frac{1}{3}\{|\vec{p_k}|^2-3\omega_{k}^2 c_s^2\}$, with $c_s$ as the velocity of sound
propagation within the medium and, $(\partial_{\mu}\mu_{a})_{P,T}=
\sum_{b=1}^{N'-1}\{\frac{\partial \mu_a}{\partial x_b}\}_{P,T,\{x_{a}\}} \partial_{\mu}x_b$, 
with $x_a$ and $\mu_a$ as the particle fraction and chemical potential associated with $a^{th}$ quantum number respectively. $q_{ak}$ is the $a^{th}$
conserved quantum number associated with $k^{th}$ component.

In order to be a solution of Eq.(\ref{RTE3}) the deviation function $\phi_{k}$ must be a linear combination of the thermodynamic forces as follows,
\begin{equation}
\phi_{k}=A_{k}X+B_{k}^{\mu}X_{q\mu}+\frac{1}{T}\sum_{a=1}^{N'-1}B_{ak}^{\mu}X_{a\mu}-C_{k}^{\mu\nu}\langle X_{\mu\nu}\rangle~,
\label{phi}
\end{equation}
where $A,B$ and $C$ are the unknown coefficients needed to be determined from the transport equation itself.
The current work majorly deals with shear viscous effects and hence from (\ref{RTE3}) and (\ref{phi}), the guiding equation to estimate shear viscosity is

\begin{equation}
 \sum_{k=1}^{N}{\cal L}_{kl}[C_{k}^{\mu\nu}]=-\frac{1}{T}f_k^0(1\pm f_k^0)\langle p_k^{\mu} p_k^{\nu}\rangle~.
 \label{GE}
\end{equation}
The unknown coefficient $C_{k}^{\mu\nu}$ is further decomposed as $C_{k}^{\mu\nu}=C_k(p_k,x)\langle{{p_k}^{\mu}}{{p_k}^{\nu}}\rangle$.
Now putting (\ref{phi}) into Eq.(\ref{Pimn}), contracting terms having compatible tensorial ranks and finally comparing with macroscopic definition of shear viscous pressure tensor
as following,
\begin{equation}
 \langle\Pi^{\mu\nu}\rangle=2\eta\langle \nabla^{\mu} u^{\nu} \rangle~,
 \label{macro}
\end{equation}
we reach the expression for shear viscosity as follows,
\begin{equation}
 \eta=-\frac{1}{10}\sum_{k=1}^{N}  \nu_k\int d\Gamma'_{p_{k}}f^{0}_k(1\pm f^{0}_k) \langle{{p_k}^{\mu}}{{p_k}^{\nu}}\rangle \langle{{p_k}_{\mu}}{{p_k}_{\nu}}\rangle C_k ~,
 \label{eta1}
\end{equation}
with $d\Gamma'_{{p}_{i}}=\frac{d^3 \vec {{p}}_{i} }{(2\pi)^3 |\vec{p}_i|}$.
The next task is to extract $C_{k}^{\mu\nu}$ from Eq.(\ref{GE}). For  this purpose, the variational approximation method has been employed where $C_{k}$ is expressed by a polynomial
of degree $p$ as follows \cite{Degroot},
\ba
C_{k}=\sum_{s=0}^{p}C_{k}^{(p)s}\tau_k^s~.
\label{Ck}
\ea

The superscript $s$ on $C_k^{(p)}$ in Eq.(\ref{Ck}) is the coefficient index belonging to each power of the polynomial expansion whereas the same on $\tau_k$ indicates 
the $s^{th}$ power of scaled energy itself. 
The coefficients $C_{k}^{(p)s}$ are functions independent of particle momenta, depending only upon particle mass and thermodynamic macroscopic
quantities. Multiplying both side of Eq.(\ref{GE}) with $\nu_k \tau_k^{r}\langle p_{k_{\mu}}p_{k_{\nu}}\rangle$ and integrating over $d\Gamma_{p_k}$ we
find the following recurrence relation,
\ba
\sum_{l,s}C_{lk}^{sr} C_{l}^{(p)s}=\nu_k \gamma_{k}^{r}~,
\label{recurrence}
\ea
with
\ba
\gamma_{k}^{r}=-\frac{1}{T}\int d\Gamma_{p_{k}}f^{0}_k(1\pm f^{0}_k) \langle p_k^{\mu}p_k^{\nu}\rangle \langle p_{k\mu}p_{k\nu}\rangle \tau_k^r~,
\label{gamma}
\ea
and
\ba
C_{lk}^{sr}=&&\nu_k\nu_l\bigg[\tau_l^s  \langle p_l^{\mu}p_l^{\nu}\rangle , \tau_k^r \langle p_{k\mu}p_{k\nu}\rangle\bigg] \nonumber\\
           +&&\nu_k \delta_{kl}\sum_m\nu_m\bigg[\tau_k^s  \langle p_k^{\mu}p_k^{\nu}\rangle , \tau_k^r \langle p_{k\mu}p_{k\nu}\rangle\bigg]\nonumber\\
           -&&\nu_k\nu_l\bigg[\tau_l^{'s}  \langle p_l^{'\mu}p_l^{'\nu}\rangle , \tau_k^r \langle p_{k\mu}p_{k\nu}\rangle\bigg] \nonumber\\
           -&&\nu_k \delta_{kl}\sum_m\nu_m\bigg[\tau_k^{'s}  \langle p_k^{'\mu}p_k^{'\nu}\rangle , \tau_k^r \langle p_{k\mu}p_{k\nu}\rangle\bigg]~.
\label{Clk}           
\ea
The bracket quantity denotes,
\ba
\bigg[A,B\bigg]=&&\int d\Gamma_{p_k}d\Gamma_{p_l}d\Gamma_{p'_l}d\Gamma_{p'_l}\nonumber\\
&&f_k^0 f_l^0 (1\pm f_k^{'0})(1\pm f_l^{'0})WAB~.
\label{bracket}
\ea
From the principle of detailed balance, it can be shown that $C_{lk}^{sr}=C_{kl}^{rs}$.

With the help of Eq.(\ref{eta1}), (\ref{Ck}) and (\ref{recurrence}) and assuming for a massless quark-gluon system the coefficients $C_{k}^{(p)}$ are species independent, 
the expression for lowest order approximation of shear viscosity is obtained as follows,
\ba
\eta=\frac{T}{10}\bigg\{\frac{\big(\sum_k\nu_k\gamma_{k}^0\big)\big(\sum_k\nu_k\gamma^{'0}_{k}\big)}{C_{11}^{11}+C_{12}^{11}+C_{21}^{11}+C_{22}^{11}}\bigg\}~,
\label{eta2}
\ea
with
\ba
\gamma_{k}^{'r}=-\frac{1}{T}\int d\Gamma'_{p_{k}}f^{0}_k(1\pm f^{0}_k) \langle p_k^{\mu}p_k^{\nu}\rangle \langle p_{k\mu}p_{k\nu}\rangle \tau_k^r~.
\label{gammap}
\ea
\subsection{Second order theory}

This section will provide the second order hydrodynamic equations for the transport quantities mentioned in the previous sections.
Before proceeding further let us summarize the thermodynamic identities, i.e, equations of motion for different first order thermodynamic state variables.
Here the Eckart's definition of velocity has been specified for the choice of reference frame. In Eckart's frame keeping terms up to second order in gradient and
ignoring the terms involving thermal and diffusive forces, 
the thermodynamic identities are given as follows,
\ba
Dn=&&-n\partial_{\mu} u^{\mu}~,
\label{Eqn}\\
D\epsilon=&&-hn\partial_{\mu} u^{\mu}+\Pi^{\mu\nu}\nabla_{\nu}u_{\mu}~,
\label{Eqep}\\
hnDu^{\mu}=&&\nabla^{\mu}P-\Delta^{\mu}_{\nu}\nabla_{\sigma}\Pi^{\nu\sigma}+(nh)^{-1}\Pi^{\mu\sigma}\nabla_{\sigma}P~,
\label{Equ}\\
DT=&&\gamma \partial_{\mu} u^{\mu}+\delta \Pi^{\mu\nu}\nabla_{\nu}u_{\mu}~,
\label{EqT}\\
D\tilde{\mu_k}=&&\gamma_k \partial_{\mu} u^{\mu}+\delta_k \Pi^{\mu\nu}\nabla_{\nu}u_{\mu}~.
\label{Eqmu}
\ea
Here, $D=u^{\mu}\partial_{\mu}$ is the covariant time derivative and $\nabla^{\mu}=\Delta^{\mu\nu}\partial_{\nu}$ is the spatial gradient. 
$h=\frac{\epsilon+P}{n}$ is the enthalpy per particle, $\tilde{\mu_k}=\frac{\mu_k}{T}$ is the scaled chemical potential and $\Pi^{\mu\nu}$ 
is the total viscous pressure tensor (including both shear and bulk part).
The coefficients in Eq.(\ref{EqT}) and (\ref{Eqmu}) are given in the appendix.

In order to obtain the second order hydrodynamic equations under EQPM model, we need to solve the relativistic transport equation (\ref{RTE1}),
keeping terms upto second order in gradient over thermodynamic quantities. For this purpose, the Grad's 14 moment method has been opted for the 
present case. In this method the relativistic transport equation picks up two additional term compared to Eq.(\ref{RTE2}) containing second order 
derivative contributions as follows,
\be
\Pi_{k}^{\mu}\partial_{\mu}f^0_{k}+f_{k}^0(1\pm f_{k}^0)\Pi_{k}^{\mu}\partial_{\mu}\phi_k+\phi_k \Pi_{k}^{\mu}\partial_{\mu}f_{k}^{0}=-\frac{1}{T}\sum_{l=1}^{N}{{\cal L}_{kl}}[\phi_{k}]~,
\label{secondhydro0}
\ee
with $\Pi_k^{\mu}=\frac{p_k^{\mu}}{T}$ as the scaled particle 4-momenta. However, the mean field force term provides zero contribution in a comoving frame.

To proceed further we need to define the deviation function $\phi_k$ for second order theory in a convenient manner. Noticing the distribution function is 
a scalar depending on the particle momentum $p_{k}^{\mu}$ and the space-time coordinate $x_{\mu}$ , the deviation function is expressed as a sum of scalar 
products of tensors formed from $p_{k}^{\mu}$ and tensor functions of $x_{\mu}$ such as,
\ba
\phi_k&&=A_k(x,\tau_k)+B_{k}^{\mu}(x,\tau_k)\langle\Pi_{k\mu}\rangle + \nonumber\\
       &&\sum_{a=1}^{N'-1}\frac{1}{T}B_{ak}^{\mu}(x,\tau_k)\langle\Pi_{k\mu}\rangle-C_{k}^{\mu\nu}(x,\tau_k)\langle\Pi_{k\mu}\Pi_{k\nu}\rangle~,
\label{Gradphi}
\ea
with the coefficients further expanded in a power series of $\tau_k$ as the following,
\ba
A_k=&&\sum_{s=0}^{2}A_{k}^{s}(x)\tau_{k}^{s}~,~~~C_{k}^{\mu\nu}=\{C_{k}^{0}(x)\}^{\mu\nu}~,\nonumber\\
B_{k}^{\mu}=&&\sum_{s=0}^{1}\{B_{k}^{s}(x)\}^{\mu}\tau_k^s~,~~~B_{ak}^{\mu}=\sum_{s=0}^{1}\{B_{ak}^{s}(x)\}^{\mu}\tau_k^s~.
\label{Gradcoeff}
\ea
Here $a$ is the index of conserved quantum number. The polynomials in Eq.(\ref{Gradcoeff}) have been retained up to the first non-vanishing contribution 
to the irreversible flows.

The next task is to express these unknown coefficients in terms of irreversible flow quantities. Here we make another assumption as before that the coefficient
functions $A_k^s, C_k^s$ etc. are species independent for a massless QGP. In the current analysis only the shear viscous flow will be discussed
ignoring the bulk viscous, thermal and diffusive parts. We first recall the traceless part of
viscous tensor under EQPM from Eq.(\ref{Pimn}). Putting (\ref{Gradphi}) into that and by the virtue of inner product property of irreducible tensors that contract only
between tensors of same rank, we obtain the coefficient $C^{0}$ as follows, 
\be
\{C^{0}\}^{\mu\nu}=\frac{5T\langle\Pi^{\mu\nu}\rangle}{\{\sum_{k}\nu_k\gamma_k^{'0}\}}~.
\label{C0}
\ee

With the help of the deviation function constructed in this way, employing the thermodynamic identities (\ref{Eqn}), (\ref{Eqep}), (\ref{Equ}), (\ref{EqT}), (\ref{Eqmu}) and
using the definition of shear viscosity from first order theory (\ref{eta2}), we finally reach the second order hydrodynamic equation
satisfied by shear viscous flow $\langle\Pi^{\alpha\beta}\rangle$ in the following way,

\ba
\langle\Pi^{\alpha\beta}\rangle=&&2\eta\langle\nabla^{\alpha}u^{\beta}\rangle-\tau_{\pi}\bigg\{\Delta^{\alpha\beta}_{\mu\nu}D\langle\Pi^{\mu\nu}\rangle\nonumber\\
            -&&2\langle\Pi_{\rho}^{\langle\alpha}\rangle\omega^{\beta\rangle\rho}+\delta_{\pi\pi}\langle\Pi^{\alpha\beta}\rangle(\partial\cdot u) \nonumber\\
            +&&\tau_{\pi\pi}\langle\Pi_{\rho}^{\langle\alpha}\rangle\langle\nabla^{\beta\rangle}u^{\rho}\rangle~\bigg\}.
\label{secondhydro}
\ea
Here, $\omega_{\mu\nu}=\frac{1}{2}(\nabla_{\mu}u_{\nu}-\nabla_{\nu}u_{\mu})$ is the vorticity tensor.
The derivation of Eq.(\ref{secondhydro}) has been detaily discussed in Appendix-A.

Clearly, we can see there are four additional terms on the right hand side of Eq.(\ref{secondhydro}) compared to first order theory (\ref{macro}). The coefficient of the second
term on the right hand side of Eq.(\ref{secondhydro}) can be identified as the relaxation time of shear viscous flow within the medium. The analytic expression of the coefficients are given as,
\ba
\tau_{\pi}=&&-10\eta T\frac{\{\sum_{k}\nu_k \gamma_k^1\}}{\{\sum_{k}\nu_k \gamma_k^0\}\{\sum_{k}\nu_k \gamma_k^{'0}\}}~,
\label{taupii}\\
\delta_{\pi\pi}=&&\frac{4}{3}\bigg[1-\frac{1}{6T^3}(1-3c_s^2)\frac{\{\sum_{k}\nu_k \sigma_k^0\}}{\{\sum_{k}\nu_k \gamma_k^1\}}\nonumber\\
                &&+\frac{c_s^2}{T^3}\frac{\{\sum_{k}\nu_k \Delta_k\delta_k^{'0}\}}{\{\sum_{k}\nu_k \gamma_k^1\}}\bigg]~,
\label{deltapipi}\\
\tau_{\pi\pi}=&&-\bigg[2+\frac{8}{21T^3}\frac{\{\sum_{k}\nu_k \sigma_k^0\}}{\{\sum_{k}\nu_k \gamma_k^1\}}\bigg]~,
\label{taupipi}
\ea
with $\gamma_k^{r}$ and $\gamma_k^{'r}$ are taken from Eq.(\ref{gamma}) and (\ref{gammap}). $\sigma_k^0$ and $\delta_k^{'0}$ are given as the follows,
\ba
\sigma_k^r=&&\int d\Gamma_{p_{k}}f_k^0(1\pm f_k^0)\mid\vec{p_k}\mid^6\tau_k^r~,\\
\delta_k^{'0}=&&\int d\Gamma'_{p_{k}}f_k^0(1\pm f_k^0)\mid\vec{p_k}\mid^5\tau_k^r~.
\ea
The relaxation time $\tau_\pi$ in Eq.(\ref{secondhydro}) are observed to be function of the first order transport coefficient $\eta$ and can be 
estimated with the help of Eq.(\ref{eta2}). The other required functions $\gamma$, $\sigma$ and $\delta$ have been estimated in Appendix-C.

\subsection{Hydro equations for a 1+1 boost invariant system}

The next task is to express the temperature evolution within the medium in the presence of viscous flows for a 1+1 boost invariant system.  For this purpose the equations
are first expressed for a system with 1 + 1 dimensional expansion in the $z$ direction. The concerned space-time variables are now the proper time $\tau$ and space-time rapidity 
$\eta_s$ which are related to the original variables ($x^{\mu}=t,z$) as $t=\tau \textrm{cosh}\eta_s$ and $z=\tau \textrm{sinh}\eta_s$. The hydrodynamic four velocity $u^{\mu}$ and 
viscous pressure tensor $\Pi^{\mu\nu}$ are then expressed as follows \cite{Muronga3},
\ba
u^{\mu}=&&(\textrm{cosh}\eta_s,0,0,\textrm{sinh} \eta_s)~,\\
\Pi^{\mu\nu}=&&\begin{bmatrix}
    \phi \textrm{sinh}^2\eta_s                &           0            &              0            &       \phi \textrm{sinh}\eta_s \textrm{cosh}\eta_s  \\
         0                                    &     -\frac{\phi}{2}    &              0            &            0                \\
         0                                    &           0            &  -\frac{\phi}{2}          &            0                \\
\phi \textrm{sinh}\eta_s \textrm{cosh}\eta_s  &           0            &              0            &      \phi \textrm{cosh}^2\eta_s  ~    \\
\end{bmatrix}~.\nonumber
\\
\ea
To perform the derivatives in a boost invariant system, we go to the Milne coordinates $(\tau,x,y,\eta_s)$ in which we find $D=\frac{\partial}{\partial \tau}$,
$\partial\cdot u=\frac{1}{\tau}$, $\langle\nabla^{\eta_s}u^{\eta_s}\rangle=-\frac{2}{3}\frac{1}{\tau^3}$ and $\phi=-\tau^2 \Pi^{\eta_s \eta_s}$. Applying these derivatives
and ignoring the bulk viscous part for the present case, Eq.(\ref{Eqep}) and (\ref{secondhydro}) turn out to be respectively,
\ba
\frac{d\epsilon}{d\tau}=&&-\frac{\epsilon+P}{\tau}+\frac{\phi}{\tau}~,
\label{hydro1}
\\
\frac{d\phi}{d\tau}=&&-\frac{\phi}{\tau_{\pi}}+\frac{2}{3}\frac{1}{\tau}\frac{2\eta}{\tau_{\pi}}-\lambda\frac{\phi}{\tau}~,
\label{hydro2}
\ea
with $\lambda=\frac{1}{3}\tau_{\pi\pi}+\delta_{\pi\pi}$.
Now, further from Eq.(\ref{Ed}) the energy density can be expressed in terms of temperature and effective chemical potential as follows,
\be
\epsilon=\frac{T^4}{\pi^2}\sum_{k}\nu_k\big\{3R_{k}^{4}+\tilde{\Delta}_kR_{k}^3\big\}~,
\label{hydro3}
\ee
with $\tilde{\Delta}_k=\frac{\Delta_k}{T}=\frac{T}{z_k}\frac{\partial z_k}{\partial T}$. The series polynomial is expressed as $R_k^n=\sum_{l}(\pm1)^{(l-1)}\frac{z_k^l}{l^n}e^{l\tilde{\mu}_{Bk}}$
for bosonic and fermionic particles respectively. $\tilde{\mu}_{Bk}=\frac{\mu_{Bk}}{T}$ is the scaled baryon chemical potential for $k^{th}$ species. With the help of Eq.(\ref{hydro3}) for a 
system with zero baryon chemical potential, (\ref{hydro1}) reduces to temperature evolution equation. Here, noticing that
$\tilde{\Delta}_k$ is a slow varying function of $T$, higher order terms like ($O\sim \tilde{\Delta}_k^2$) have been ignored keeping upto linear terms only.
Following this prescription, the final second order hydrodynamic equations for temperature and viscous flow come out to be,
\ba
\frac{dT}{d\tau}=&&-a_1\frac{T}{\tau}+\frac{b_1}{T^3}\frac{\phi}{\tau}~,
\label{hydro6}\\
\frac{d\phi}{d\tau}=&&-\frac{\phi}{\tau_\pi}+\frac{4}{3}\frac{a_2}{\tau}-b_2\frac{\phi}{\tau}~.
\label{hydro7}
\ea
The corresponding coefficients are described as follows,
\ba
a_1=&&(1+c_s^2)\frac{\sum_{k}\nu_k\big\{3R_k^4+\tilde{\Delta}_kR_k^3\big\}}{\big\{12\sum_{k}\nu_kR_k^4+8\sum_{k}\nu_k\tilde{\Delta}_kR_k^3\big\}}~,\\
b_1=&&\frac{\pi^2}{\big\{12\sum_{k}\nu_kR_k^4+8\sum_{k}\nu_k\tilde{\Delta}_kR_k^3\big\}}~,\\
a_2=&&\frac{\eta}{\tau_\pi}~,
\label{hydro4}\\
b_2=&&\lambda=\frac{1}{3}\tau_{\pi\pi}+\delta_{\pi\pi}~.
\label{hydro5}
\ea
The analytic expressions of $\gamma_k^r$, $\gamma_k^{'r}$, $\sigma_k^0$ and $\delta_k^{'0}$ functions upon which $\tau_\pi$, $\delta_{\pi\pi}$ ans $\tau_{\pi\pi}$ crucially depend, are 
given in the appendix. The values of sound velocity and entropy density are given by the relationships, 
$c_s^2=\sum_{k}\big\{\frac{\partial P_k}{\partial T}+\frac{\partial P_k}{\partial \mu_{Bk}}\tilde{\mu}_{Bk}\big\}/\sum_{k}\big\{\frac{\partial \epsilon_k}{\partial T}+\frac{\partial 
\epsilon_k}{\partial \mu_{Bk}}\tilde{\mu}_{Bk}\big\}$ 
and $s=\frac{\sum_{k}\{\epsilon_k+P_k\}}{T}-\frac{\sum_k\{n_k\mu_{Bk}\}}{T}$ respectively.

\section{Results}
Eq.(\ref{hydro6}) and (\ref{hydro7}) describe the evolution of temperature and viscous pressure tensor within a dissipative QCD medium described by EQPM model.
In this section I will present the numerical solution of these two equations for a 1+1 boost invariant system in order to visualize the effect of EQPM model
on the temperature and viscous pressure evolution. The fugacity parameters $z_{g,q}$ have been set from the updated lattice EOSs \cite{Bazavov} for the current
purpose and in order to do so the transition temperature $T_c$ has been considered to be 170 MeV throughout the analysis. 

Before  proceeding 
further, we need to examine the behavior of the input parameters of the hydrodynamic
evolution of these coupled equations, namely $\tau_{\pi}$, $\delta_{\pi\pi}$ and $\tau_{\pi\pi}$ from Eq.(\ref{taupii}), (\ref{deltapipi}) and (\ref{taupipi}) respectively.
At Israel-Stewart (IS) limit (with $z_g/z_q\rightarrow 1$) they take the values as $\tau_{\pi}^{IS}=\frac{3}{2}\frac{\eta}{P}$, $\delta_{\pi\pi}^{IS}=\frac{4}{3}$ and $\tau_{\pi\pi}^{IS}=2$, 
which agree with \cite{Muronga1,Jaiswal,Romatschke,Denicol0}. The next task is to show their behavior including the equation of state effects through the EQPM model.

\begin{figure}[ht]
\includegraphics[scale=0.35]{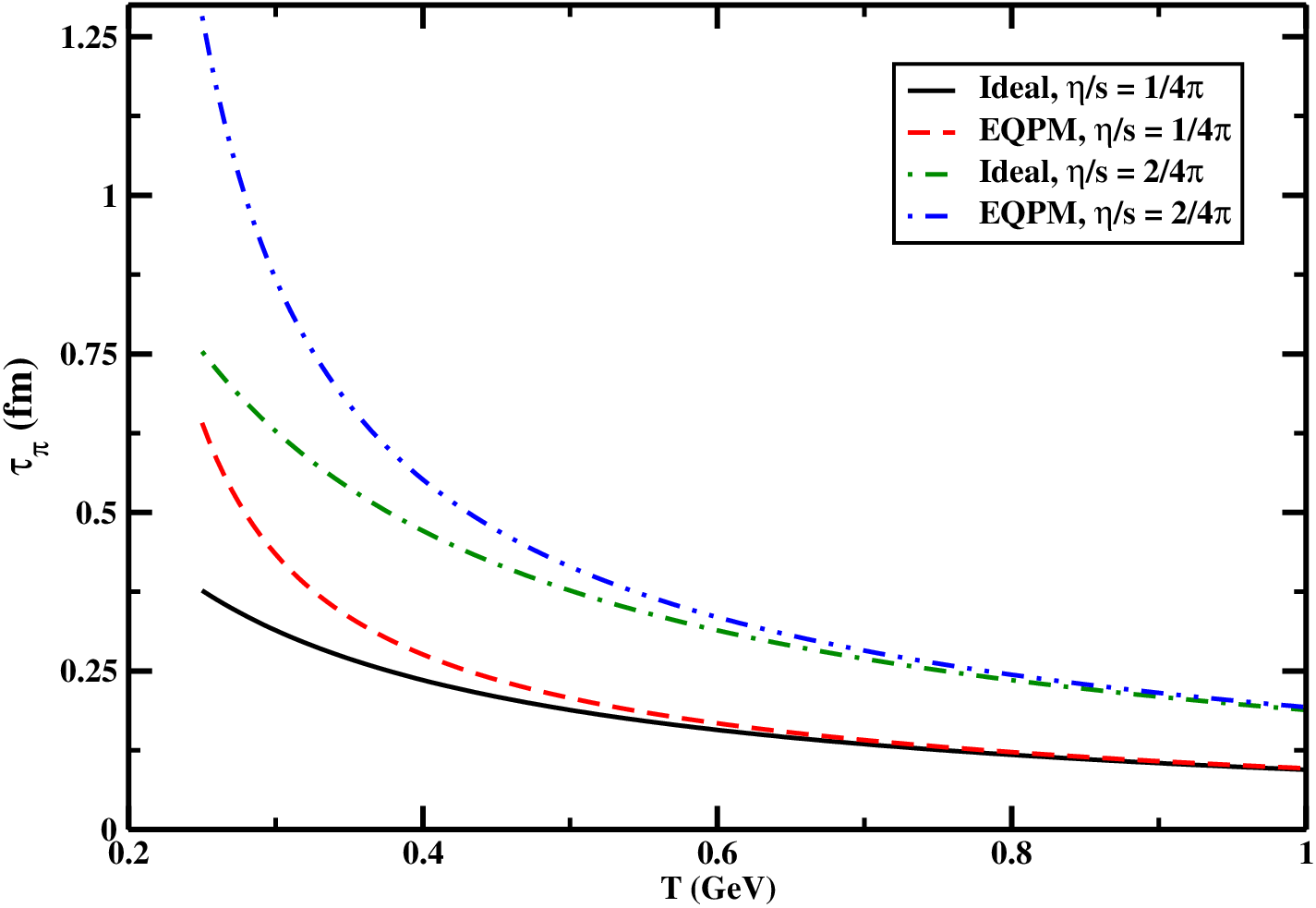}
\caption{Relaxation time of shear viscous flow as a function of temperature for different $\eta/s$ with and without EQPM.}
\label{taupi}
\end{figure}

\begin{figure}[ht]
\includegraphics[scale=0.35]{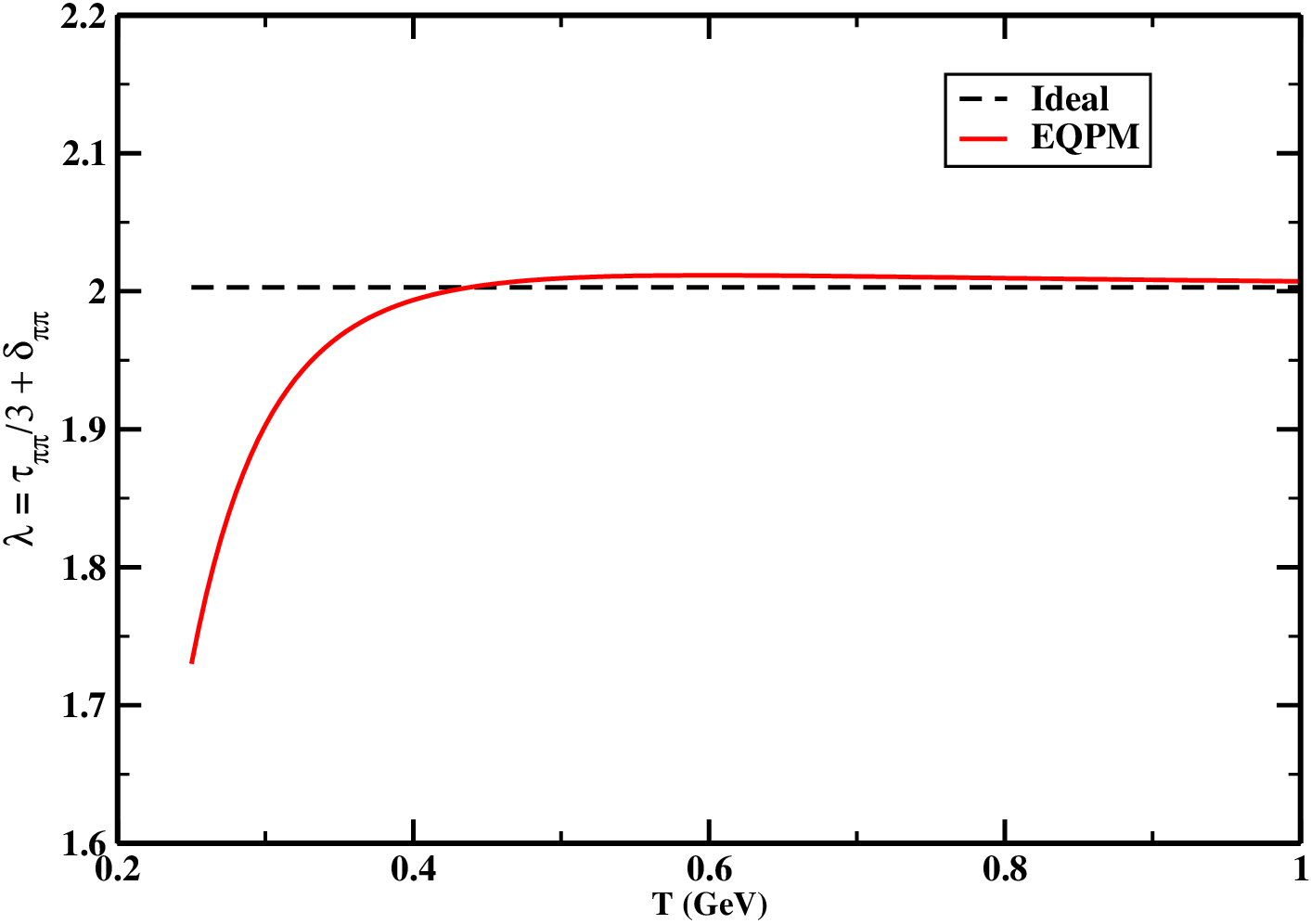}
\caption{$\lambda$ as a function of temperature with and without EQPM.}
\label{lambda}
\end{figure}

\begin{figure}[ht]
\includegraphics[scale=0.35]{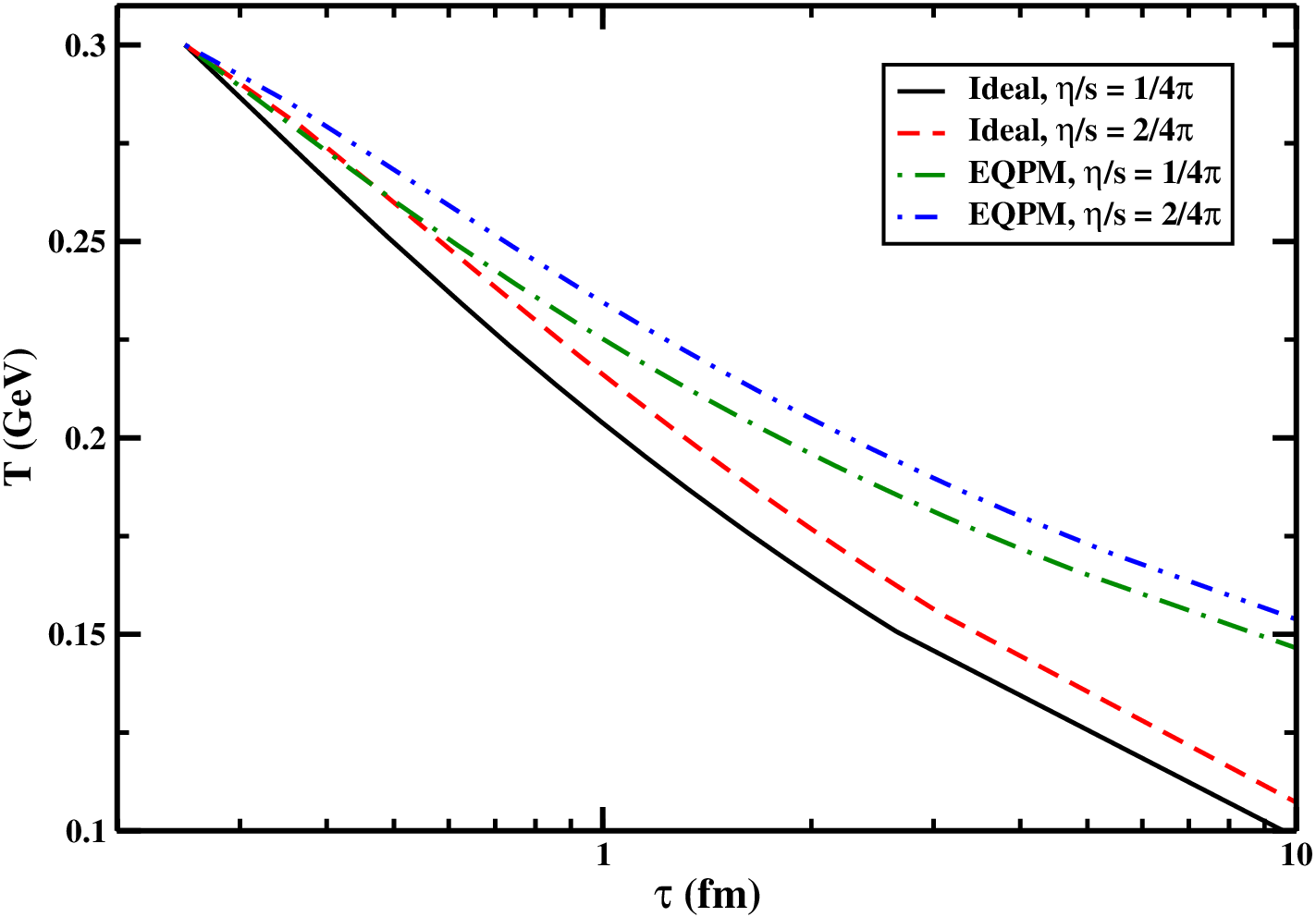}
\caption{Proper time evolution of temperature for different $\eta/s$ with and without EQPM.}
\label{Ttau}
\end{figure}

\begin{figure}[ht]
\includegraphics[scale=0.35]{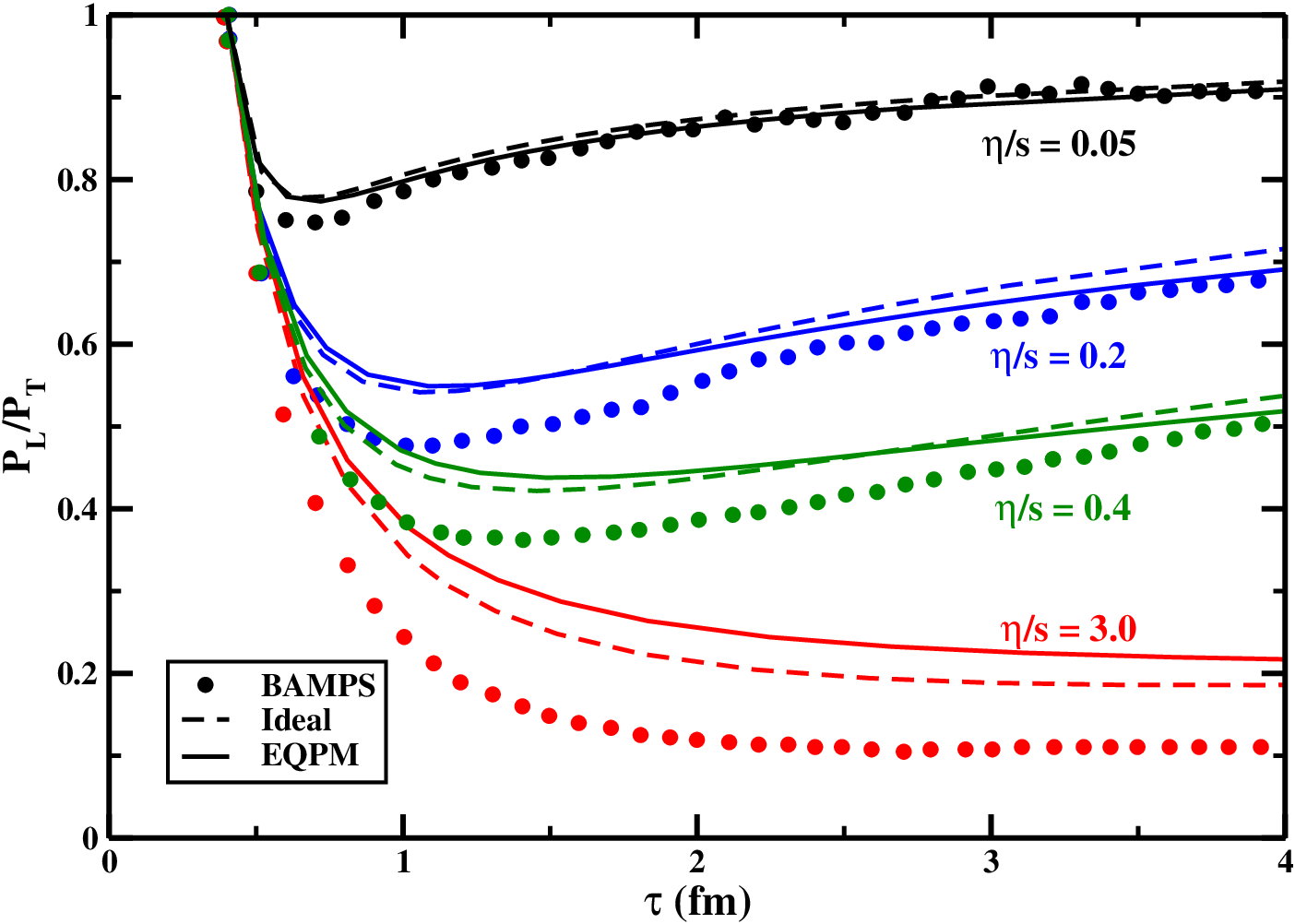}
\caption{Proper time evolution of $P_L/P_T$ for different $\eta/s$ with and without EQPM.}
\label{PLPT1}
\end{figure}

\begin{figure}[ht]
\includegraphics[scale=0.35]{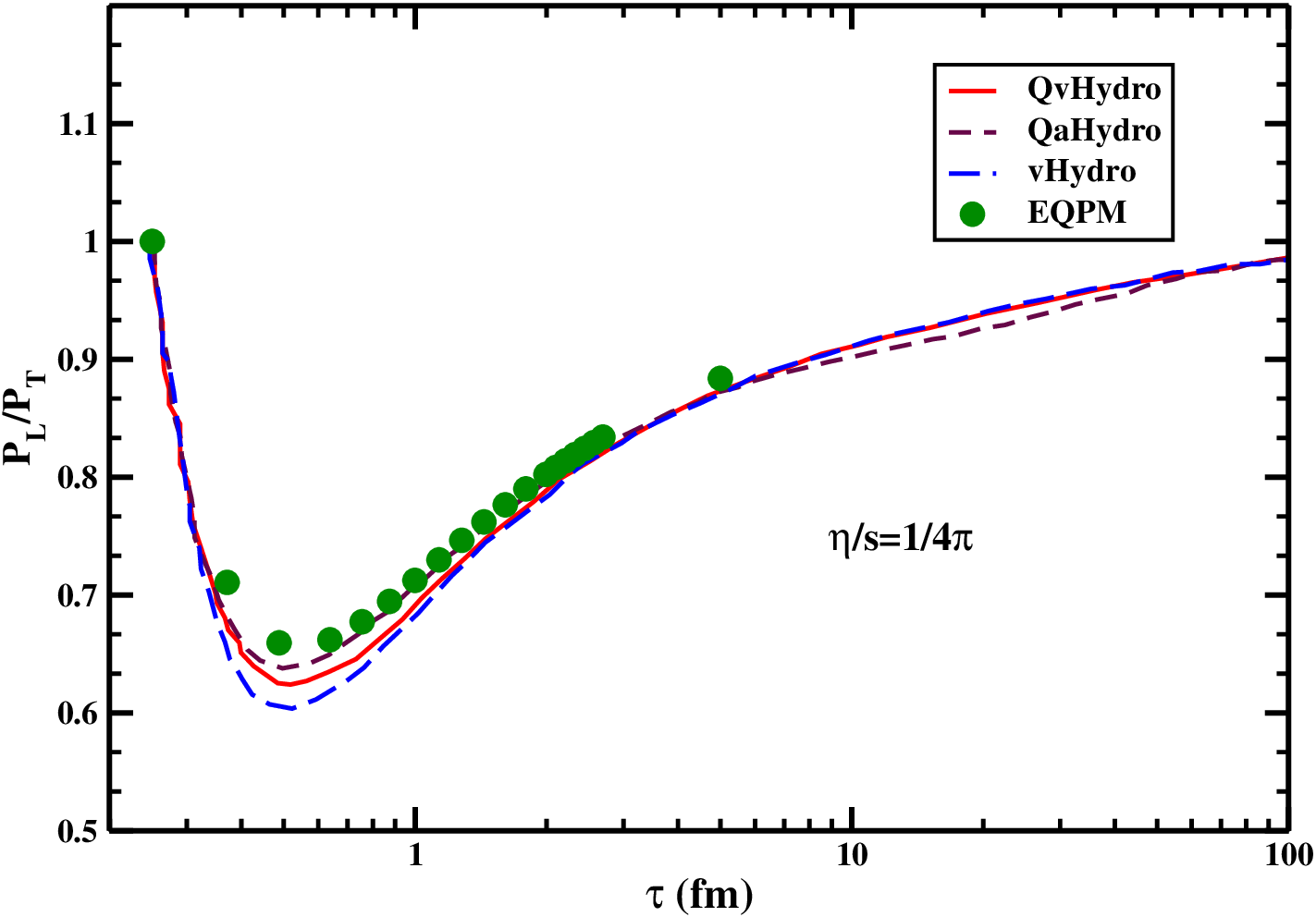}
\caption{Proper time evolution of $P_L/P_T$ and comparing with different quasiparticle models.}
\label{PLPT2}
\end{figure}

Fig. (\ref{taupi}) shows the temperature dependence of $\tau_{\pi}$ with and without EQPM model for different values of shear viscosity. As predicted by
\cite{Okamura,Denicol,Koide} the temperature dependence of $\tau_{\pi}$ shows an decreasing trend with increasing temperature and being proportional to
shear viscosity shows higher value for higher $\eta/s$. For each set of $\eta/s$, $\tau_\pi$ has been plotted with and without considering the EQPM 
model taken into calculation. In high temperature domain  the plots with and without EQPM model merge with each other indicating that at those temperatures, 
the quasi-particle properties behave almost like those of the free particles, with $z_k$'s reaching their Stefan-Boltzmann (SB) limit, $z_g/z_q\rightarrow 1$.
However, at low temperatures the effect of EQPM model is quite prominent with the enhancement over ideal values (without EQPM model). 
\\
\\
\\
In Fig.(\ref{lambda}) the temperature behavior of $\lambda=\frac{1}{3}\tau_{\pi\pi}+\delta_{\pi\pi}$ has been depicted with and without EQPM model. 
In ideal case, i.e, in SB limit $\lambda$ turns out to be $2$ which has been indicated by the dashed line. For the EQPM case (solid line), the ratio 
deviates largely from $2$ at low temperature regions due to the equation of state effects, whereas at high temperatures it predictably tends to merge with the ideal result. 
\\
\\
Next, the proper time evolution of the temperature has been shown in Fig.(\ref{Ttau}) for different set of $\eta/s$ ratio with and without EQPM model.
The initial values of proper time and temperature have been taken to be the Relativistic Heavy Ion Collider (RHIC) values \cite{El}, $\tau_i=0.25 fm$ and $T_i=0.3 GeV$. 
The initial value of shear viscous pressure is taken
to be the Navier-Stokes value $\phi_i=\frac{4\eta}{3\tau_i}$. For higher value of viscosity the curves are enhanced indicating dissipative processes
make the system to take larger times to cool down. The effect of EQPM model for each set of $\eta/s$ is clearly distinct from the ideal ones which
is more prominent at larger times i.e. for smaller temperatures.
\\
\\
\\
Next, in Fig.(\ref{PLPT1}) and (\ref{PLPT2}) proper time evolution of the pressure anisotropy, i.e the ratio between longitudinal and  transverse pressure
$P_L/P_T=(P-\phi)/(P+\frac{\phi}{2})$ has been depicted using EQPM model. In Fig.(\ref{PLPT1}), the EQPM results have been shown along with ideal
Israel-Stewart (IS) results and BAMPS data \cite{BAMPS2} for different
$\eta/s$ values with initial temperature $T_i=0.5 GeV$ and initial time $\tau_i=0.4fm$, which correspond to Large Hadron Collider (LHC) initial condition \cite{El}.
The EQPM results are clearly distinct from ideal (i.e, Israel-Stewart) results which is more significant for larger dissipation, i.e,  greater $\eta/s$. 
It is observed that the massless IS theory is much closer to the BAMPS results than EQPM one, which does not show close agreement with BAMPS. This is expected because
of the very different equation of state between BAMPS which uses ultrarelativistic EOS ($P=\frac{1}{3}\epsilon$) and the current quasiparticle model which uses
lattice EOS.

\begin{figure}[ht]
\includegraphics[scale=0.35]{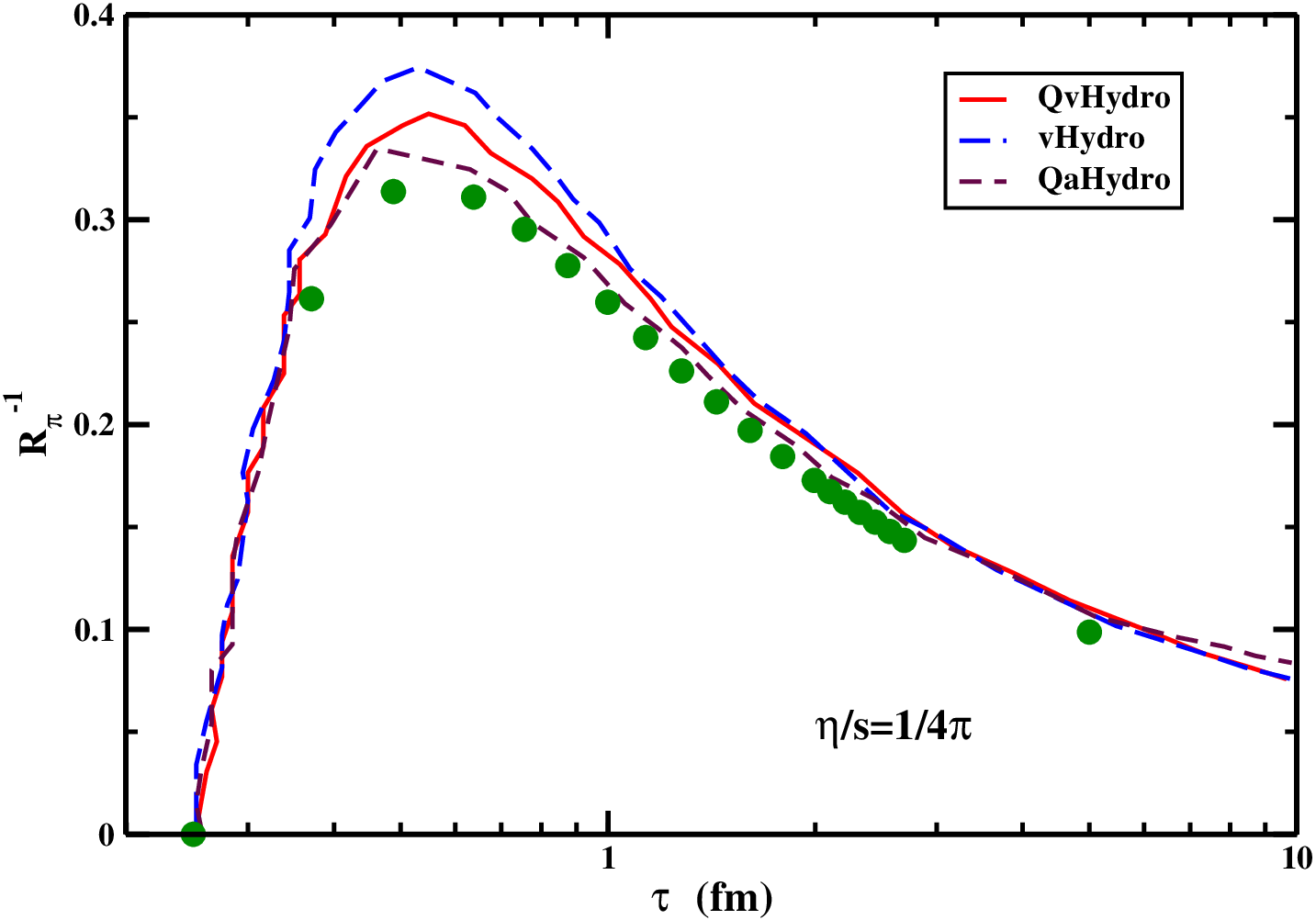}
\caption{Proper time evolution of inverse Reynolds number.}
\label{Reynolds}
\end{figure}

In Fig.(\ref{PLPT2}) the same has
been plotted with $T_i=0.6 GeV$, $\tau_i=0.25fm$ (which are again LHC conditions) and $\eta/s=1/4\pi$. In both the cases (Fig.(\ref{PLPT1}) and Fig.(\ref{PLPT2})) the initial pressure configuration has been taken to be isotropic $\phi_i=0$.
In Fig.(\ref{PLPT2}) the EQPM result of pressure anisotropy has been compared with other quasi particle models. vHydro stands for the standard second-order
viscous hydrodynamics, where QaHydro is the quasiparticle anisotropic hydrodynamics \cite{Alqahtani} and QvHydro is quasiparticle second-order viscous hydrodynamics\cite{Tinti}.
These two quasi particle models describe the thermodynamic system by considering temperature dependent quasiparticle masses. The prediction of pressure anisotropy
from current EQPM set up appears to be consistent with these models as well as showing small quantitative difference because of the separate approaches adopted in them. 
\\
\\
\\
\\
\\
\\
The last plot presents proper time evolution of the inverse Reynolds number $R_{\pi}^{-1}=\sqrt{\frac{3}{2}}\{\phi/P\}$ in Fig.({\ref{Reynolds}}) with the same initial 
conditions as Fig.(\ref{PLPT2}) and is compared with other quasiparticle estimates discussed earlier. Here also EQPM is observed to predict consistent result with other models.
\\
\\

\section{Conclusion and Outlook}
In the present work a second order relativistic viscous hydrodynamic theory has been developed including the effects of a strongly interacting thermal medium
through EQPM model. The effects of hot QCD EOSs have been included through the temperature dependent fugacity parameter of the quasi-partons which have been
extracted from an updated lattice prediction. The second order hydro equations have been formulated from the covariant kinetic theory for a multi-component
system that has been consistently developed within the scheme of EQPM model. The relativistic transport equation from this effective kinetic theory has been
solved using Grad's 14 moment method within a viscous medium. Finally, the hydrodynamic equations corresponding to energy density (as well as temperature)
and viscous flow have been solved for a 1+1 boost invariant system. The proper time evolution of temperature and pressure anisotropy have shown significant
modification due to the inclusion of EQPM model with respect to the ideal ones. Hence, it can be concluded that the effects of a strongly interacting
thermal QCD medium have been embedded into the evolution equations of the system through the EQPM model which is reflected on the behavior of thermodynamic 
quantities and cooling laws of the system consequently.

Formulating second order dissipative hydrodynamics in the presence of electromagnetic fields is becoming an interesting and relevant topic now a days.
Works like \cite{Denicol2,Hernandez,Gursoy,Kharzeev,Roy,Inghirami} are worth mentioning and encouraging in this direction.
Keeping this scenario in view, an immediate extension of the current work will be treating the system with EQPM model in the presence of a strong electromagnetic field. A number of recent works
regarding the first order transport processes in the presence of a strong magnetic field employing EQPM model have been pursued in \cite{Kurian}. Based on
these previous works a complete formalism of dissipative magnetohydrodynamic theory within the scope of EQPM model is a sure aim in the future. 

\appendix

\section{Derivation of hydro equation}
In this section the details of the derivation of Eq.(\ref{secondhydro}) from (\ref{secondhydro0}) has been discussed. 
First the quantity $\Pi_{k}^{\mu}\partial_{\mu}f_k^0$ is decomposed in the following way with the help of Eq.(\ref{EqT}) and (\ref{Eqmu}),
\ba
\Pi_{k}^{\mu}\partial_{\mu}f_k^0=&&f_k^{0}(1\pm f_k^0)\times\nonumber\\
&&\bigg[\big\{ \hat{Q}_k\partial\cdot u +\hat{Q}^{'}_k (\Pi^{\mu\nu}\nabla_{\nu}u_{\mu}) \big\}\nonumber\\
&&+\Pi_{k}^{\mu}\big\{\tau_k\frac{\nabla_{\mu}T}{T}+\nabla_{\mu}\tilde{\mu}_k-\tau_kDu_{\mu}\big\}\nonumber\\
&&-\langle\Pi_k^{\mu}\Pi_k^{\nu}\rangle \langle \nabla_{\mu}u_{\nu}\rangle\bigg]~,
\label{iden}
\ea
with $\hat{Q}_k=\frac{1}{3}\{\frac{\mid\vec{p_k}\mid^2}{T^2}-3c_s^2\tau_k^2\}=\frac{1}{3}\frac{\mid\vec{p_k}\mid^2}{T^2}\{1-3c_s^2(1+2\frac{\Delta_k}{\mid\vec{p_k}\mid})\}$ and $\hat{Q}_k^{'}=\frac{\tau_k^2}{T}\delta+\tau_k\delta_k$.

In order to obtain Eq.(\ref{secondhydro}), we multiply both sides of Eq.(\ref{secondhydro0})
with $\nu_k\langle\Pi_k^{\alpha}\Pi_k^{\beta}\rangle$, integrate over $d\Gamma_{p_k}$ and finally sum over $k$. Now let us investigate the resulting equation term by term. 
Before poceeding further let us review the moment integral of the particle distribution in the following manner,
\ba
F_k^{\nu_1\nu_2\cdots\nu_n}=\int d\Gamma_{p_k}f_k^0(1\pm f_k^0)p_k^{\nu_1\nu_2\cdots\nu_n}~,
\ea
which can be given in terms of hydrodynamic velocity and projection operator as below,
\ba
F_k^{\nu_1\nu_2\cdots\nu_n}=\sum_{l=0}^{[n/2]}a_{nl}(\Delta u)_{nl}~,
\ea
with 
\be
a_{nl}=\frac{(-1)^l}{2\pi^2} \frac{^{n}C_{2l}}{2l+1}\int\frac{d\mid\vec{p_k}\mid}{\omega_{p_k}}\mid\vec{p_k}\mid^{(2l+2)}\omega_{p_k}^{n-2l}f_k^0(1\pm f_k^0)~,\nonumber\\
\ee
and
\be
(\Delta u_{nl})=\frac{1}{n!}\sum_{P}\Delta^{\nu_1\nu_2}\cdots\Delta^{\nu_{2l-1}\nu_{2l}}u^{\nu_{2l+1}}\cdots u^{\nu_n}~,
\ee
where the summation is extended over all permutations P of the indices.

Applying this definition it can be shown that,
\ba
&&\int d\Gamma_{p_k}f_k^0 (1\pm f_k^0)\tau_k \langle\Pi_k^{\alpha}\Pi_k^{\beta}\rangle=0~,\\
&&\int d\Gamma_{p_k}f_k^0 (1\pm f_k^0)\tau_k^2 \langle\Pi_k^{\alpha}\Pi_k^{\beta}\rangle=0~,\\
&&\int d\Gamma_{p_k}f_k^0 (1\pm f_k^0)\langle\Pi_k^{\mu}\rangle\langle\Pi_k^{\alpha}\Pi_k^{\beta}\rangle=0~,\\
&&\int d\Gamma_{p_k}f_k^0 (1\pm f_k^0)\langle\Pi_k^{\mu}\rangle\tau_k\langle\Pi_k^{\alpha}\Pi_k^{\beta}\rangle=0~,\\
&&\int d\Gamma_{p_k}f_k^0 (1\pm f_k^0)\langle\Pi_k^{\mu}\rangle\langle\Pi_{k\mu}\rangle\langle\Pi_k^{\alpha}\Pi_k^{\beta}\rangle=0~,\\
\ea

which establishes the property of irreducible tensors that the inner product of two irreducible tensors with different rank must be zero. Then using these identities the first term on left hand side 
of (\ref{secondhydro0}) straightforwardly comes out to be,
\be
\textrm{[LHS]}_{\textrm{I}}=\frac{\langle\nabla^{\alpha}u^{\beta}\rangle}{5T^3}\big\{\sum_{k}\nu_k\gamma_{k}^{0}\big\}~.
\label{LHS1}
\ee
Here I have used the following inner product property,
\ba
&&\int d\Gamma_{p_k}f_k^0(1\pm f_k^0)\langle\Pi_k^{\mu}\Pi_k^{\nu}\rangle \langle\Pi_k^{\alpha}\Pi_k^{\beta}\rangle \nonumber\\=
&&\int d\Gamma_{p_k}f_k^0(1\pm f_k^0)\langle\Pi_k^{\mu}\Pi_k^{\nu}\rangle \langle\Pi_{k\mu}\Pi_{k\nu}\rangle\frac{1}{5}\Delta^{\mu\nu\alpha\beta}~,
\ea
by the virtue of the complete contraction property of traceless projection tensor $\Delta_{(n)}^{(n)}=2n+1$ with $n$ being the respective rank \cite{Degroot}.

However, the second term on the left hand side contains the derivative over the deviation function $\phi_k$. Decomposing the derivative in a temporal and spatial part such as
$\partial^{\mu}=u^{\mu}D+\nabla^{\mu}$, putting the expression of $\phi_k$ from Eq.(\ref{Gradphi}) and using the contraction property of irreducible tensors,
we finally reach at,
\ba
\textrm{[LHS]}_{\textrm{II}}&&=\frac{1}{T^2}\frac{\big\{\sum_{k}\nu_k\gamma_k^1\big\}}{\big\{\sum_k\nu_k\gamma_k^{'0}\big\}}\bigg[\Delta^{\alpha\beta}_{\mu\nu}D\langle\Pi^{\mu\nu}\rangle\nonumber\\
&&+\frac{4}{3} \langle\Pi^{\alpha\beta}\rangle (\partial\cdot u)-2\langle\Pi_{\rho}^{\langle\alpha}\rangle\omega^{\beta\rangle\rho}\nonumber\\
&&-2\langle\Pi_{\rho}^{\langle\alpha}\rangle\langle\nabla^{\beta\rangle}u^{\rho}\rangle\bigg]~.
\label{LHS22}
\ea
The third term contains the product of Eq.(\ref{Gradphi}) and (\ref{iden}). After the tensor contraction and ignoring the bulk viscous, thermal and diffusive effects it takes the following form,
\ba
\textrm{[LHS]}_{\textrm{III}}=&&-\frac{5T}{\{\sum_{k}\nu_k\gamma_k^{'0}\}}(\partial\cdot u)\langle{\Pi}_{\mu\nu} \rangle\nonumber\\
&&\times\bigg\{\sum_{k}\nu_k\int dF_{p_k} \hat{Q}_k \langle\Pi_k^{\mu}\Pi_k^{\nu}\rangle \langle\Pi_k^{\alpha}\Pi_k^{\beta}\rangle\bigg\}\nonumber\\
&&+\frac{5T}{\{\sum_k\nu_k\gamma_k^{'0}\}}\langle\Pi_{\mu\nu}\rangle\langle\nabla_{\mu_1}u_{\nu_1}\rangle\nonumber\\
&&\times\bigg\{\sum_{k}\nu_k\int dF_{p_k} \langle\Pi_k^{\mu}\Pi_k^{\nu}\rangle \langle\Pi_k^{\mu_1}\Pi_k^{\nu_1}\rangle\langle\Pi_k^{\alpha}\Pi_k^{\beta}\rangle\bigg\}\nonumber\\
\label{iden1}
\ea
with $dF_{p_k}=d\Gamma_{p_k}f_k^0(1\pm f_k^0)$.
After pursuing the moment integrals it becomes,
\ba
\textrm{[LHS]}_{\textrm{III}}&&=-\frac{2}{9T^5} (\partial\cdot u)\langle{\Pi}^{\alpha\beta}\rangle   
\bigg\{(1-3c_s^2)\frac{\{\sum_{k}\nu_k \sigma_k^0\}}{\{\sum_{k}\nu_k \gamma_k^{'0}\}}\nonumber\\
&&-6 c_s^2\frac{\{\sum_{k}\nu_k \Delta_k\delta_k^{'0}\}}{\{\sum_{k}\nu_k \gamma_k^{'0}\}}  \bigg\} \nonumber\\
&&-\frac{8}{21T^5}\frac{\{\sum_{k}\nu_k \sigma_k^0\}}{\{\sum_{k}\nu_k \gamma_k^{'0}\}}\langle\Pi_{\rho}^{\langle\alpha}\rangle\langle\nabla^{\beta\rangle}u^{\rho}\rangle~.
\label{LHS3}
\ea

Lastly, we focus the term resulting from the right hand side of Eq.(\ref{secondhydro0}). Again using the definition of $\phi_k$ from Eq.(\ref{Gradphi}), applying
the inner product property of irreducible tensor and employing the definition of $C_{lk}^{sr}$ from Eq.(\ref{Clk}) we come to the conclusion,
\be
\textrm{[RHS]}=\frac{1}{5T^5}\{C^{0}\}^{\alpha\beta}\sum_{k,l}C_{lk}^{00}~.
\ee
with the help of the expression of shear viscosity from Eq.(\ref{eta2}) and employing Eq.(\ref{C0}) we finally reach at,
\be
\textrm{[RHS]}=\frac{1}{10T^3}\frac{1}{\eta}\big\{\sum_k\nu_k\gamma_k^0\big\}\langle\Pi^{\alpha\beta}\rangle~.
\label{RHS}
\ee
With the help of Eq.(\ref{LHS1}), (\ref{LHS22}), (\ref{LHS3}) and (\ref{RHS}), we finally reach at Eq.(\ref{secondhydro}). 

\section{Important identities}
\begin{widetext}
The coefficients in Eq.(\ref{EqT}) and (\ref{Eqmu}) are given for a two component system as the following,

\ba
\gamma=&&\frac{-nh\big(\frac{\partial n_1}{\partial \tilde{\mu_1}}\big)\big(\frac{\partial n_2}{\partial \tilde{\mu_2}}\big)
             +n_1\big(\frac{\partial \epsilon_1}{\partial \tilde{\mu_1}}\big)\big(\frac{\partial n_2}{\partial \tilde{\mu_2}}\big)
             +n_2\big(\frac{\partial \epsilon_2}{\partial \tilde{\mu_2}}\big)\big(\frac{\partial n_1}{\partial \tilde{\mu_1}}\big)}
              {\big(\frac{\partial\epsilon}{\partial T}\big)\big(\frac{\partial n_1}{\partial \tilde{\mu_1}}\big)\big(\frac{\partial n_2}{\partial \tilde{\mu_2}}\big)
            -\big(\frac{\partial\epsilon_1}{\partial \tilde{\mu_1}}\big)\big(\frac{\partial n_1}{\partial T}\big)\big(\frac{\partial n_2}{\partial \tilde{\mu_2}}\big)
            -\big(\frac{\partial\epsilon_2}{\partial \tilde{\mu_2}}\big)\big(\frac{\partial n_2}{\partial T}\big)\big(\frac{\partial n_1}{\partial \tilde{\mu_1}}\big) }~,
\label{gama}           
\\
\delta=&&\frac{\big(\frac{\partial n_1}{\partial \tilde{\mu_2}}\big)\big(\frac{\partial n_2}{\partial \tilde{\mu_2}}\big)}
              {\big(\frac{\partial\epsilon}{\partial T}\big)\big(\frac{\partial n_1}{\partial \tilde{\mu_1}}\big)\big(\frac{\partial n_2}{\partial \tilde{\mu_2}}\big)
            -\big(\frac{\partial\epsilon_1}{\partial \tilde{\mu_1}}\big)\big(\frac{\partial n_1}{\partial T}\big)\big(\frac{\partial n_2}{\partial \tilde{\mu_2}}\big)
            -\big(\frac{\partial\epsilon_2}{\partial \tilde{\mu_2}}\big)\big(\frac{\partial n_2}{\partial T}\big)\big(\frac{\partial n_1}{\partial \tilde{\mu_1}}\big) }~,
\label{delta}
\\
\gamma_1=&&\frac{-n_1\big(\frac{\partial \epsilon}{\partial T}\big)\big(\frac{\partial n_2}{\partial \tilde{\mu_2}}\big)
             +n_1\big(\frac{\partial \epsilon_2}{\partial \tilde{\mu_2}}\big)\big(\frac{\partial n_2}{\partial T}\big)
             -n_2\big(\frac{\partial \epsilon_2}{\partial \tilde{\mu_2}}\big)\big(\frac{\partial n_1}{\partial T}\big)
             +nh\big(\frac{\partial n_1}{\partial T}\big)\big(\frac{\partial n_2}{\partial \tilde{\mu_2}}\big)}
              {\big(\frac{\partial\epsilon}{\partial T}\big)\big(\frac{\partial n_1}{\partial \tilde{\mu_1}}\big)\big(\frac{\partial n_2}{\partial \tilde{\mu_2}}\big)
            -\big(\frac{\partial\epsilon_1}{\partial \tilde{\mu_1}}\big)\big(\frac{\partial n_1}{\partial T}\big)\big(\frac{\partial n_2}{\partial \tilde{\mu_2}}\big)
            -\big(\frac{\partial\epsilon_2}{\partial \tilde{\mu_2}}\big)\big(\frac{\partial n_2}{\partial T}\big)\big(\frac{\partial n_1}{\partial \tilde{\mu_1}}\big) }~,
            \label{gama1}
\\            
\delta_1=&&\frac{-\big(\frac{\partial n_1}{\partial T}\big)\big(\frac{\partial n_2}{\partial \tilde{\mu_2}}\big)}
              {\big(\frac{\partial\epsilon}{\partial T}\big)\big(\frac{\partial n_1}{\partial \tilde{\mu_1}}\big)\big(\frac{\partial n_2}{\partial \tilde{\mu_2}}\big)
            -\big(\frac{\partial\epsilon_1}{\partial \tilde{\mu_1}}\big)\big(\frac{\partial n_1}{\partial T}\big)\big(\frac{\partial n_2}{\partial \tilde{\mu_2}}\big)
            -\big(\frac{\partial\epsilon_2}{\partial \tilde{\mu_2}}\big)\big(\frac{\partial n_2}{\partial T}\big)\big(\frac{\partial n_1}{\partial \tilde{\mu_1}}\big) }~.
            \label{delta1}
\ea

The expression for $\gamma_2$ and $\delta_2$ can be trivially obtained by mutually interchanging particle index 1 and 2 in Eq.(\ref{gama1}) and (\ref{delta1}) respectively. 


\end{widetext}

\section{Analytic expression for $\gamma_k^r$ functions}

\ba
\gamma_k^0=&&-\frac{1}{3\pi^2 T}\big\{I_k^5-\Delta_k I_k^4\big\}~,\\
\gamma_k^1=&&-\frac{1}{3\pi^2 T^2}I_k^6 ~,\\
\gamma_k^2=&&-\frac{1}{3\pi^2 T^3}\big\{I_k^7+\Delta_k I_k^6\big\}~,\\
\gamma_k^{'0}=&&-\frac{1}{3\pi^2 T}I_k^5 ~,\\
\sigma_{k}^0=&&\frac{1}{2\pi^2}\{I_k^7-\Delta_k I_k^6\}~,\\
\delta_k^{'0}=&&\frac{1}{2\pi^2}I_k^6~.
\ea
with $I_k^n=n!T^{n+1}R_k^n$ where $R_k^n=\sum_{l=1}^{N}(\pm)^{l-1}\frac{z_k^l}{l^n}$ for Boson and Fermion respectively in absence of baryon chemical potential.

\acknowledgments
I duly acknowledge the Science and Engineering Research Board (SERB) and Indo-U.S. Science and Technology Forum (IUSSTF) for funding the Indo-US postdoctoral fellowship 
and providing the financial support.

\end{document}